\documentclass[amssymb,prb,twocolumn,floats]{revtex4}
\usepackage{bm}
\usepackage{graphicx}
\begin{document}
\preprint{April. 27, 2001}
\title{Magnetic domains in III-V magnetic semiconductors}
\author{T. Dietl}
\email{dietl@ifpan.edu.pl}
\homepage{http://www.ifpan.edu.pl/SL-2/sl23.html}
\affiliation{Institute of Physics and College of Science, Polish
Academy of Sciences,  al. Lotnik\'{o}w 32/46, PL-00-668 Warszawa,
Poland}
\author{J\"urgen K\"onig}
\affiliation{Institut f\"ur Theoretische Festk\"orperphysik,
Universit\"at Karlsruhe, 76128 Karlsruhe, Germany\\
Department of Physics, The University of Texas,
Austin, TX 78712}
\author{A.H. MacDonald}
\affiliation{Department of Physics, The University of Texas,
Austin, TX 78712}
\date{\today}

\begin{abstract}
Recent progress in theoretical understanding of magnetic anisotropy
and stiffness in III-V magnetic semiconductors is exploited for
predictions of magnetic domain characteristics and methods of their
tuning. We evaluate the width and the energy of domain walls as well
as the period of stripe domains in perpendicular films. The
computed stripe width $d = 1.1$ $\mu$m for
Ga$_{0.957}$Mn$_{0.043}$As/In$_{0.16}$Ga$_{0.84}$As compares
favorably to the experimental value 1.5 $\mu$m, as determined by
Shono {\it et al.} [{\it Appl. Phys. Lett.} {\bf 77}, 1363 (2000)].
\end{abstract}

\maketitle


The utility foreseen for ferromagnetic semiconductors rests on the
possibility of tailoring their electronic and magnetic properties
on the same footing. Indeed, successful control of the Curie
temperature $T_{\mbox{\small {C}}}$ by hole density has been
achieved in Mn-based IV-VI,\cite{Stor86} III-V,\cite{Oiwa97} and
II-VI\cite{Haur97,Ferr01} semiconductor compounds. It has also been
found that both biaxial strain imposed by lattice
mismatch\cite{Ohno96b} and confinement in quantum
structures\cite{Haur97} can serve to engineer the direction of the
easy axis. Furthermore, light-induced changes of the magnetic phase
in\cite{Kosh97} (In,Mn)As/(Al,Ga)Sb and\cite{Haur97}
(Cd,Mn)Te/(Cd,Zn,Mg)Te heterostructures has been demonstrated.
Finally, the possibility to tune $T_{\mbox{\small {C}}}$ of
(In,Mn)As quantum wells by metallic gates has also been
shown.\cite{Ohno00} We can anticipate many new demonstrations of
tunable magnetic and magnetotransport properties in the future.

These tuning capabilities along with $T_{\mbox{\small {C}}}$'s as high
as\cite{Ohno96a} 110 K in (Ga,Mn)As with 5.3\% Mn have
triggered a considerable theoretical effort to elucidate
the origin of ferromagnetism in III-V magnetic semiconductors.
While there is general agreement that the Mn constituent
introduces both localized spins and itinerant holes, the
nature and energy of the Mn-derived states, the role of intrinsic
defects, the relative importance of charge and spin fluctuations,
as well as the consequences of electrostatic and magnetic disorder
are still under debate.\cite{Diet01a}

Recently, a quantitative theory of hole-mediated ferromagnetism in
tetrahedrally coordinated magnetic semiconductors has been put
forward by the independent effort of two teams consisting of the
present authors and co-workers.\cite{Diet00,Abol01,Diet01b,Koni01}
In this theory, the ferromagnetic interaction between spins
localized on the d shells of the magnetic ions is mediated by holes
in the valence band. The free energy of the hole liquid is computed
by diagonalizing the 6x6 Luttinger Hamiltonian, which contains
$k\cdot p$, spin-orbit, and p-d exchange interactions, the latter
taken into account in the molecular-field and virtual-crystal
approximations. The influence of electrostatic and spin disorder on
magnetic properties is neglected, since they are not expected to
have a qualitative impact on spin-polarized band electron
thermodynamic properties. Hole-hole interactions can be taken into
account in the spirit of Fermi liquid theory. The use of a
mean-field approximation for the coupled band-electron and
local-moment systems can be justified by the long-range character
of the carrier-mediated spin-spin interaction, at least when the
ratio of hole to Mn density is small and the Fermi energy is
large.\cite{Schl01} It has been shown\cite{Diet01b} that this
model, with material parameters known from independent experiments,
satisfactorily explains the magnitude of $T_{\mbox{\small{C}}}$,
the temperature dependence of the spontaneous magnetization, the
strength and strain dependence of the magnetic anisotropy, as well
as the spectral dependence of the magnetic circular dichroism in
(Ga,Mn)As. The hole-spin polarization has also been
evaluated.\cite{Diet01b} To address spin fluctuations a theoretical
description of ferromagnetism beyond the mean-field approximation
has been developed,\cite{Koni00} from which the magnon excitation
spectrum,\cite{Koni00,Koni01} and the magnetic stiffness could be
deduced.\cite{Koni01}

In this paper, we exploit this progress in the theoretical
description of magnetic
anisotropy\cite{Diet00,Abol01,Diet01b,Koni01} and magnetic
stiffness\cite{Koni01} to address the domain structure in epitaxial
layers of (Ga,Mn)As. The comparison between the computed and
experimentally observed width of domain stripes presented here
constitutes an additional test of current theory.  Moreover, the
determined values for anisotropy energy, domain-wall energy and
width may serve for optimizing the design of, for instance, spin
injection,\cite{Ohno99b} spin tunneling,\cite{Tana01,Chib00}, or
micromechanical\cite{Harr99} nanostructures of (Ga,Mn)As.

For the calculation presented below we adopt band-structure
parameters, elastic constants, and deformation potentials of GaAs,
a set of values employed in our previous
works\cite{Diet00,Abol01,Diet01b,Koni01} for (Ga,Mn)As.
The Mn spins are assumed to be in the d$^5$ configuration, so
that $S=5/2$ and the Mn Land\'e factor $g=2.0$.
For the p-d exchange energy we take\cite{Diet00,Diet01b}
$\beta N_o = -1.2$ eV, which for the cation concentration of GaAs,
$N_o = 2.21\times10^{22}$ cm$^{-3}$, corresponds to
$J_{\rm pd}\equiv -\beta = 0.054$ eVnm$^3$. The Fermi
liquid parameter $A_F = 1.2$ enters the enhancement of
$T_{\mbox{\small {C}}}$ and of the p-d exchange splitting $\Delta
=A_FJ_{\rm pd} M/(g\mu_B)$ of the
valence band at magnetization $M$ of the Mn spins.\cite{Diet01b}

Another important parameter characterizing epitaxial layers is the
magnitude of biaxial strain. It depends on the layer thickness $d$
and the difference between the lattice parameters of the substrate
and the layer, $\Delta a = a_s - a(x)$. In the case of (Ga,Mn)As
that is obtained by low-temperature MBE, films with $d$ as large as
2 $\mu$m are not relaxed.\cite{Shen99} For such layers, the
relevant components of the strain tensor assume the form
$\epsilon_{xx} = \epsilon_{yy}= \Delta a/a$ and $\epsilon_{zz} =
-2\epsilon_{xx}c_{12}/c_{11}$, where the $c_{ij}$ are elastic
constants. Since $\mbox{d}a(x)/\mbox{d} x = 0.032$
nm,\cite{Ohno96b} the appropriately thin layer of (Ga,Mn)As
deposited on GaAs or (Al,Ga)As is under compressive strain but the
use of (Ga,In)As substrates can result in a tensile strain.

It has been demonstrated by Ohno {\it et al.}\cite{Ohno96b} that
for films under compressive (tensile) strain the easy axis is in
plane (perpendicular to film growth direction). Quantitatively,
$\epsilon_{xx} = -0.2$\% for the Ga$_{0.0965}$Mn$_{0.035}$As film
on GaAs, for which Ohno {\it et al.}\cite{Ohno96b} determined the
magnitude of the magnetic field $H_a$ aligning magnetization along
the hard axis. Tensile strain of $\epsilon_{xx} = 0.90$\% is
expected for the 0.2 $\mu$m perpendicular film of
Ga$_{0.957}$Mn$_{0.043}$As on Ga$_{0.84}$In$_{0.16}$As, a sample
employed by Shono {\it et al.}\cite{Shon00} to examine the domain
structure. Importantly, the theory referred to
above\cite{Diet00,Abol01,Diet01b} reproduces correctly the strain
dependence of magnetic anisotropy and, in particular, explains the
magnitude of $H_a$.\cite{Diet01b} Furthermore, it has been
found\cite{Abol01,Diet01b} that for the relevant values of strain,
hole concentrations $p$, and magnitudes of $M$, the energy density
$K_u$ that characterizes uniaxial magnetic anisotropy is greater
than the corresponding cubic anisotropy terms $K_c$ and the energy
density of the stray fields $K_d = \mu_oM^2/2$. Thus, (Ga,Mn)As can
be classified as a uniaxial ferromagnet.

We discuss the domain structure in terms of micromagnetic
theory.\cite{Hube98,Skom99} Within this approach, a uniaxial
ferromagnet is characterized by the anisotropy energy $K_u$, the
magnetic stiffness $A$, and the saturation magnetization $M_s$. The
procedure we adopt here consists of evaluating $M(T)$, and thus
$\Delta(T)$, in mean-field approximation. Then, $K_u(\Delta)$ and
$A(\Delta)$ are calculated. We believe that this procedure is well
grounded at low temperatures. However fluctuation corrections will
be more important at non-zero $T$, particularly in the critical
region near $T_{\mbox{\small {C}}}$.

An important question arises whether the continuous-medium
approximation underlying the micromagnetic theory is valid in
diluted magnetic semiconductors, which contain a relatively low
concentration of magnetic ions and even lower concentration of
carriers. To address this question we note that the shortest length
scale of micromagnetic theory is the width of the Bloch domain
wall,
\begin{equation}
\delta_W = \pi \sqrt{A/K_u}.
\end{equation}
This length has to be compared to the mean distance between holes,
$r_m = \sqrt[3]{3/(4\pi p)}$. For concreteness, we consider
Ga$_{0.957}$Mn$_{0.043}$As on Ga$_{0.84}$In$_{0.16}$As for which
$T_{\mbox{\small {C}}} = 80$ K.\cite{Shon00} Assuming $p=3\times
10^{20}$ cm$^{-3}$ we obtain the mean-field value of
$T_{\mbox{\small {C}}} = 91$ K. The actual value of $p$ is
uncertain; the mean-field $T_{\mbox{\small {C}}}$ would be 80 K for
$p=2.5\times 10^{20}$ cm$^{-3}$. Since the Fermi energy is greater
than $\Delta$ at $T\rightarrow 0$ (this corresponds to the
``weak-coupling'' or ``RKKY'' regime,\cite{Koni01b}) the
redistribution of holes between the four valence subbands is only
partial, and both $A(\Delta)$ and $K_u(\Delta)$ are proportional to
$\Delta^2$, except for large $\Delta$, where the increase of $K_u$
with $\Delta$ is somewhat weaker. Hence, $A/K_u$ and thus
$\delta_W$ are virtually independent of $\Delta$, that is of $T$.
For instance, for $p = 3\times 10^{20}$ cm$^{-3}$, $\delta_W =
14.9$ nm if $T/T_{\mbox{\small {C}}} \rightarrow 0$ and decreases
to 14.6 nm in the opposite limit $T/T_{\mbox{\small {C}}}
\rightarrow 1$. Thus, the value determined for $\delta_W \approx
15$ nm is by more than a factor of ten longer than $r_m$. We
checked that $\delta_W \gg r_m$ in the whole relevant range of the
hole concentrations down to $10^{20}$ cm$^{-3}$. We conclude that
the micromagnetic theory in its standard continuous-medium form is
suitable for modeling the domain structure in (Ga,Mn)As.

In the case of perpendicular easy-axis films, a competition between
energies associated with the stray fields and the formation of the
Bloch domain walls results in a simple stripe domain structure in
the demagnetized thermal equilibrium state.\cite{Kooy60} Indeed,
Shono {\it et al.} observed such a structure in
Ga$_{0.957}$Mn$_{0.043}$As/Ga$_{0.84}$In$_{0.16}$As by means of a
micro-Hall scanning probe.\cite{Shon00} In particular, well defined
stripes of width $W$ increasing from 1.5 $\mu$m  at 9 K to 2.5
$\mu$m at 30 K were observed. Above 60 K, the stripes were less
regular. Their width was evaluated to change from 3 to 6 $\mu$m in
the temperature range between 65 and 77 K. The stripes were
oriented along [110] at low temperatures but tended to lay along
[100] above 60 K.

To interpret the experimental results we recall\cite{Hube98,Kooy60}
that for $d \gg \delta_W$ and $K_u \gg K_d$, the stripe width $W$ is
determined by a solution of the transcendental equation,
\begin{equation}
\lambda_c = (P^2/\pi^3) \sum_{n=1,3,5,..} n^{-3}[1- (1 + 2\pi
n/P)e^{-2\pi n/P}].
\end{equation}
Here, $P$ is the normalized stripe period $P=2W/d$ and the
parameter $\lambda_c$ describes the ratio of the Bloch domain-wall
energy $\gamma_W = 4\sqrt{AK_u}$ and the stray-field energy $K_d$,
\begin{equation}
\lambda_c = 4\sqrt{AK_u}/(\mu_oM^2d).
\end{equation}
Figure 1 shows $\lambda_c$ as a function of $T/T_{\mbox{\small
{C}}}$ computed for the film in question. Again the dependence on
$T$ is weak as both the denominator and numerator are to a good
accuracy proportional to $\Delta^2$.

\begin{figure}
\includegraphics*[width=90mm]{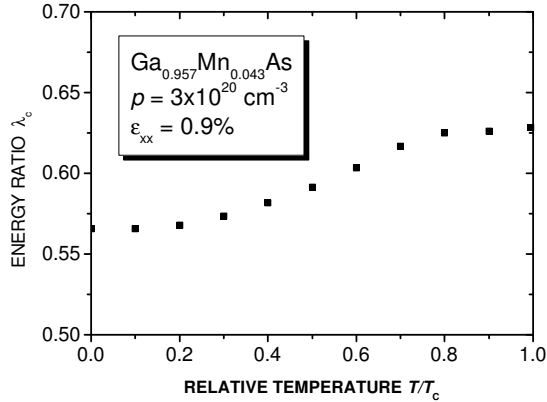}
\caption[]{Computed ratio of domain wall to stray-field energy,
$\lambda_c$ (Eq.~5) as a function of reduced temperature for a
Ga$_{1-x}$Mn$_{x}$As film with the easy axis along the growth
direction.} \label{fig:RC_1}
\end{figure}

\begin{figure}
\includegraphics*[width=90mm]{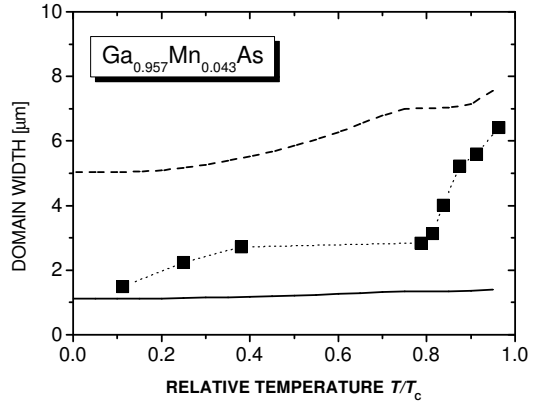}
\caption[]{Temperature dependence of the width of domain stripes as
measured by Shono {\it et al.}\cite{Shon00} for the
Ga$_{0.957}$Mn$_{0.043}$As film with the easy axis along the growth
direction (full squares). Computed domain width for the parameter
$\lambda_c$ depicted in Fig.~\ref{fig:RC_1} is shown by the solid
line. The dashed line is computed assuming that $\lambda_c$ is by a
factor of 1.8 greater.}
 \label{fig:RC_2}
\end{figure}

We now evaluate the domain width $W(T)$. The results and a
comparison to the experimental data of Shono {\it et
al.}\cite{Shon00} are presented in Fig.~2. Furthermore, in order to
establish the sensitivity of the theoretical results to the
parameter values, we include results calculated for a value of
$\lambda_c$ 1.8 times larger as well. We see that the computed
value for low temperatures, $W = 1.1$ $\mu$m, compares favorably
with the experimental finding, $W = 1.5$ $\mu$m. The difference
between the two may stem, in principle, from the uncertainty in the
input parameters. We note, however, that recent Monte Carlo
simulations\cite{Schl01b} demonstrates that disorder in positions
of magnetic ions tends to enhance $A$, and thus $W$. At the same
time, it is clear from Fig.~\ref{fig:RC_2} that our model predicts
much weaker temperature dependence of $W$ than observed
experimentally. It seems likely that the break in the experimental
dependence $W(T)$ around 60 K marks the beginning of the critical
regime. In this regime, long-length-scale fluctuations in $M$, not
accounted for in our theory, become important. These will have a
larger effect on $K_d$, which is sensitive to fluctuations on
scales shorter than $W$, than on the domain-wall energy $\gamma_W$,
which is sensitive only to fluctuations on length scales shorter
than $\delta_W$.

Next, we comment on the crystallographic orientation of the domain
stripes. The computed direction of the in-plane easy axis
oscillates between [100] and [110] directions as a function of the
hole concentration $p$ and the splitting
$\Delta$.\cite{Abol01,Diet01b} For the film in question we find
that the lowest energy will have domain walls laying along [100] or
equivalent directions, provided that $p$ is greater than $2.5\times
10^{20}$ cm$^{-3}$. This stripe direction is in disagreement with
the experimental results\cite{Shon00} described above. However, for
slightly smaller values of $p$, the predicted behavior is in accord
with the findings. In particular, for $p=2\times 10^{20}$
cm$^{-3}$, the wall direction is expected to assume [110] or
equivalent directions for large $\Delta \rightarrow \Delta_s$, that
is at low temperatures, but to extend along [100] or equivalent
directions at higher $T$, $\Delta<\Delta_s/2$.

It is well known that the domain structure of perpendicular
easy-axis films exhibits an interesting evolution as a function of
the magnetic field along the easy axis. When the field increases,
stripes magnetized along the field grow, while those antiparallel
to the field shrink. However, beyond a critical field value,
cylindrical bubbles rather than stripes have lower energy. It would
be interesting to search for such domains in perpendicular films of
III-V magnetic semiconductors. At the same time, history-dependent
metastable domain arrangements are expected to develop, for
instance, a ``froth'' structure in the remanent state. The magnetic
field $H_{cb}$ at which domains vanish entirely under conditions of
thermal equilibrium increases with $1/\lambda_c$, so that $H_{cb}
\rightarrow M_s$ for $\lambda_c \ll 1$. For the case under
consideration, $\lambda_c \approx 0.5$ according to
Fig.~\ref{fig:RC_1}, which corresponds to $\mu_oH_{cb} \approx
0.1\mu_oM_s \approx 5.5$ mT. This is consistent with the observed
coercive force $\mu_oH_c \approx 20$ mT and square hysteresis for
such a film.\cite{Shen99} Actually, the fact that $H_{c} > H_{cb}$
implies the existence of a domain pinning mechanism.

In view of the interest in magnetic nanocrystals, the length scale
below which a ferromagnetic sample is in a single-domain state, is
an interesting material parameter. Such a length scale is shape
dependent. We consider square samples of ferromagnetic films with
dimensions $W\times W \times d$. In thermodynamic equilibrium, the
width $W_{SD}$ below which the material is in a single-domain state
is determined by $d$ and $l_c = \gamma_W/2K_d =\lambda_cd$. For the
perpendicular film discussed above, $d =0.2$ $\mu$m and $l_c = 0.1$
$\mu$m. For these values, according to numerical results of Hubert
and Sch\"afer,\cite{Hube98} $W_{SD}=1.2$ $\mu$m. For such small
single-domain particles, the celebrated Stoner-Wohlfarth theory
predicts an abrupt switching of the magnetization direction in a
magnetic field along the easy axis at $H_f= H_a \equiv
2K_u/(\mu_oM_s)$. Thus, in this case the coercive force is equal to
the anisotropy field aligning magnetization along the hard axis,
$\mu_oH_c =\mu_oH_a =  670$ mT for the material in question. It
would be interesting to check experimentally the actual magnitude
of $H_c$ in nanostructures of III-V DMS. On the other hand, it is
known already that in the case of macroscopic films $H_c \ll H_a$,
reflecting the inevitability of domain nucleation processes,
associated--for instance--with space fluctuations of $H_a$. The
appearance of reverse domains followed by domain wall motion
results in the complete reversal of magnetization at $H_c$ in the
range $H_{a} > H_c > H_{cb}$. The wall motion begins when the
field-induced torque on the wall magnetization overcomes the wall
pinning force.

In conclusion, our results imply that despite relatively small
concentrations of magnetic ions and carriers, domain properties of
III-V DMS are, in many respects, similar to those of standard
ferromagnets. In particular, domain characteristic can be described
in terms of micromagnetic theory. Such an approach, combined with
our microscopic theory of hole-mediated ferromagnetism, predicts
the direction and strength of magnetic anisotropy as well as
characteristic dimensions of the domains correctly. We note that
the values predicted for both anisotropy energies and domain-wall
energies are qualitatively dependent on an accurate representation
of the host semiconductor valence band. Further experimental
studies on both macroscopic films and nanostructures with fine
spatial and time resolution will certainly improve our
understanding of this novel ferromagnetic system, opening the doors
wider for domain engineering. Furthermore, the role of the
magnetostriction and, in particular, its contribution to the strain
tensor remains to be elucidated. On the theoretical side, it will
be interesting to see to what extent intrinsic fluctuations in the
Mn distribution, the distribution of other extrinsic defects, and
the carrier density distributions they produce, account for domain
nucleation and pinning fields.

\section*{Acknowledgments}
We thank T. Jungwirth, F. Matsukura, H. Ohno, and J. Schliemann for
valuable discussions. The work was supported by the Foundation for
Polish Science, the Deutsche Forschungsgemeinschaft, the Indiana
21st Century Fund, the DARPA spintronics program, and the Welch
foundation.

\end{document}